\documentclass[aps,prb,twocolumn,superscriptaddress,showpacs,10pt]{revtex4-1}
\usepackage{graphicx}
\usepackage{calc}
\usepackage{bm}
\usepackage{color}

\begin{document}

\title{Conductance oscillations at the interface between a superconductor and the helical edge channel in a narrow HgTe quantum well.}

\author{A.~Kononov}
\affiliation{Institute of Solid State Physics RAS, 142432 Chernogolovka, Russia}
\author{S.V.~Egorov}
\affiliation{Institute of Solid State Physics RAS, 142432 Chernogolovka, Russia}
\author{N.~Titova}
\affiliation{Moscow State Pedagogical University, Moscow 119991, Russia}
\author{Z. D. Kvon}
\affiliation{Institute of Semiconductor Physics, Novosibirsk 630090, Russia}
\author{N. N. Mikhailov}
\affiliation{Institute of Semiconductor Physics, Novosibirsk 630090, Russia}
\author{S. A. Dvoretsky}
\affiliation{Institute of Semiconductor Physics, Novosibirsk 630090, Russia}
\author{E.V.~Deviatov}
\affiliation{Institute of Solid State Physics RAS, 142432 Chernogolovka, Russia}
\affiliation{Moscow Institute of Physics and Technology, Dolgoprudny 141700 Russia}

\date{\today}

\begin{abstract}
We  experimentally investigate  electron transport through the interface between a superconductor and the edge of a two-dimensional electron system with band inversion. The interface is realized as a tunnel NbN side contact to a narrow 8~nm HgTe quantum well. It demonstrates a typical Andreev behavior with finite conductance within the superconducting gap. Surprisingly, the conductance is modulated by a number of equally-spaced oscillations. The oscillations are present only within the superconducting gap and at lowest, below 1~K, temperatures. The oscillations disappear completely in magnetic fields, normal to the two-dimensional electron system plane. In contrast,  the oscillations' period is only weakly affected by the highest, up to 14~T, in-plane oriented magnetic fields. We interpret this behavior as the interference oscillations in a helical one-dimensional edge channel due to a proximity with a superconductor.
\end{abstract}


\maketitle


Recently, a strong interest appears to the investigations of electron transport through the interface between a superconductor (S) and a normal (N) semiconductor-based low-dimensional structure. Because of the Majorana problem~\cite{reviews}, this interest is mostly connected with different SNS type structures, where the N region is a one-dimensional quantum wire~\cite{Lutchyn,Oreg,Pientka,Heiblum,Mourik,Deng} or a topological surface state~\cite{Fu,yakoby}. 

An attractive example of a one-dimensional topological surface state is a current-carrying helical edge channel, realized in a narrow HgTe quantum well~\cite{konig}. This channel originates if the well thickness exceeds the critical 6.3~nm due to the inverted band structure in the bulk HgTe two-dimensional system~\cite{pankratov,zhang1,kane,zhang2}. In contrast to the conventional quantum Hall edge states~\cite{buttiker}, this channel is helical, i.e. it consists of two spin-resolved counter-propagating states in zero magnetic field. Despite the initial  idea of a topological protection~\cite{konig,zhang1,kane,zhang2}, backscattering appears at macroscopic distances~\cite{kvon,kvon_nonlocal}, possibly due to the allowed two-particle process~\cite{mirlin} and to the electron puddles~\cite{glazman}. The edge current has been directly demonstrated in a visualization experiment~\cite{imaging} in zero magnetic field. Also, the supercurrents have been investigated for a two-terminal SNS Josephson junction, with a rectangular section of quantum well located between two superconducting leads~\cite{yakoby}. So, a narrow HgTe quantum well is a  promising candidate~\cite{yakoby_top} for the search for a topological superconductivity~\cite{Sau1,Potter}.

On the other hand, even a single SN interface is predicted to demonstrate a number of intriguing effects, e.g. conductance oscillations due to a proximity effect~\cite{chevallier,adroguer}, giant spin rotation~\cite{tanaka09}, and localized edge states~\cite{guigou}. Pronounced Fabry-Perot oscillations have been demonstrated for a three-dimensional $Bi_2Se_3$ topological insulator sandwiched between a superconducting and normal leads~\cite{finck}. In this experiment, a proximity with a superconductor doubled the period of the oscillations, although they were present also for normal leads. Thus, it seems to be reasonable to investigate electron transport in a single SN side contact at the edge of a  narrow HgTe quantum well. 

Here, we experimentally investigate  electron transport through the interface between a superconductor and the edge of a two-dimensional electron system with band inversion. The interface is realized as a tunnel NbN side contact to a narrow 8~nm HgTe quantum well. It demonstrates a typical Andreev behavior with finite conductance within the superconducting gap. Surprisingly, the conductance is modulated by a number of equally-spaced oscillations. The oscillations are present only within the superconducting gap and at lowest, below 1~K, temperatures. The oscillations disappear completely in magnetic fields, normal to the two-dimensional electron system plane. In contrast,  the oscillations' period is only weakly affected by the highest, up to 14~T, in-plane oriented magnetic fields. We interpret this behavior as the predicted~\cite{adroguer} interference oscillations in a helical one-dimensional edge channel due to a proximity with a superconductor.


\begin{figure}
\includegraphics[width=\columnwidth]{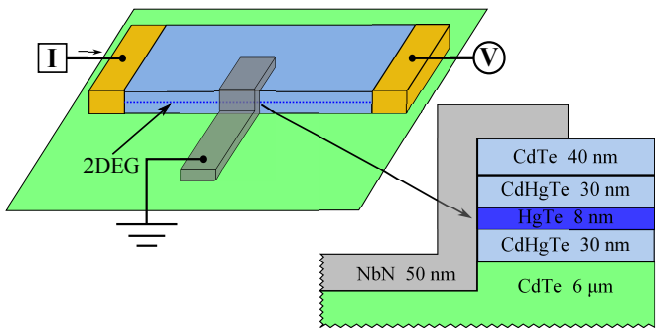}
\caption{(Color online) Sketch of the sample (not in scale) with electrical connections. The 100~$\mu$m wide mesa has two In Ohmic contacts (yellow). The 20~$\mu$m wide superconducting NbN  stripe (gray) is placed at the mesa step, with low (2-3~$\mu$m) overlap. Because of the insulating layer on the top of the structure, a {\em side} SN junction is formed  between the NbN electrode and the 2DEG edge at the mesa step. We study electron transport across one particular NbN-2DEG junction in a three-point configuration: a current is applied between one of the In Ohmic contacts and a superconducting NbN electrode which is grounded while the other In contact measures the 2DEG potential. Because of the relatively low in-plane  2DEG resistance, and the low resistance of the superconducting NbN electrode, the measured $dV/dI(V)$ curves reflect the behavior of the NbN-2DEG interface.}
\label{sample}
\end{figure}

Our $Cd_{0.65}Hg_{0.35}Te/HgTe/Cd_{0.65}Hg_{0.35}Te$ quantum well with [013] surface orientation and width $d=8$~nm is grown by molecular beam epitaxy. A detailed description of the well structure is given elsewhere~\cite{growth1,growth2}. Because $d$ exceeds the critical value 6.3~nm, the quantum well is characterized by band inversion~\cite{kvon,kvon_nonlocal}. It contains a two-dimensional electron gas (2DEG) with the electron  density $1.5 \cdot 10^{11}  $cm$^{-2}$ and the low-temperature mobility $2\cdot 10^{5}  $cm$^{2}$/Vs, as obtained from standard magnetoresistance measurements.

A sample sketch is presented in Fig.~\ref{sample}. A 200 nm high mesa step is formed by dry etching in Ar plasma. We fabricate two Ohmic contacts to the 2DEG by annealing In.  In addition, we use dc magnetron sputtering to deposit a 50~nm thick superconducting NbN  film  at the mesa step, the surface is mildly cleaned by Ar plasma before sputtering. To avoid any 2DEG  degradation, the sample is not  heated during the sputtering process. The 20~$\mu$m wide NbN stripe is formed by lift-off technique, with low (2-3~$\mu$m) mesa overlap, see Fig.~\ref{sample}.

Because of the insulating layer on the top of the structure, a {\em side} SN junction is formed  between the NbN electrode and the 2DEG edge at the mesa step. In samples with etched mesa an insulating region of finite width is always present at the 2DEG edge~\cite{shklovskii,image02}. In our samples this region is significant enough to provide  tunnel S-2DEG junctions, which are characterized by $0.5 - 1.5$~M$\Omega$ normal resistances $R_N$.

\begin{figure}
\includegraphics[width=0.9\columnwidth]{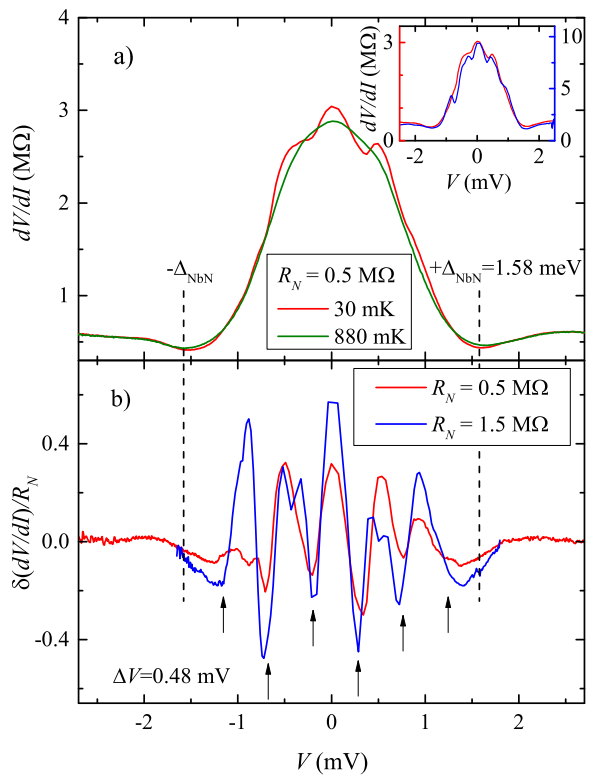}
\caption{(Color online) (a) Examples of  $dV/dI(V)$ characteristics for a single NbN-2DEG side junction in zero magnetic field at two different temperatures $T<<T_c=11$~K.  The $dV/dI(V)$ curves are of standard Andreev behavior~\protect\cite{tinkham}. They demonstrate a clearly visible NbN superconducting gap $\Delta_{NbN}=\pm$~1.58~mV (denoted by dashed lines), which corresponds well to the directly measured  $T_c=11$~K. Within the gap at  $|eV|<\Delta_{NbN}$, the maximum differential resistance $R_{max}$ is undoubtedly finite. Inset demonstrates a perfect scaling of two low-temperature (30~mK) $dV/dI(V)$ curves for the junctions with different normal resistances $R_N$ (0.5~M$\Omega$ and 1.5~M$\Omega$ respectively), i.e. a constant resistance ratio $R_N/R_{max}\sim 0.16$. The pronounced $dV/dI$ oscillations can be clearly seen at  $|eV|<\Delta_{NbN}$ for the lowest (30~mK) temperature. The oscillating behavior is suppressed completely above 0.88~K, while the $dV/dI(V)$ curve itself is not affected at $T<<T_c$.
(b) Normalized $dV/dI(V)/R_N$ oscillations, demonstrated in detail by subtracting the high-temperature (0.88~K) monotonous curve from the low-temperature (30~mK) oscillating one. The result is shown for two junctions with strongly different $R_N$. The positions of the conductance oscillations are denoted by arrows. They coincide well for different $R_N$ and correspond to a constant period  $\Delta V=0.48$~mV. The oscillations are concentrated strictly within the superconducting gap. 
}
\label{IV}
\end{figure}

We study electron transport across one particular NbN-2DEG junction in a three-point configuration: a current is applied between one of the In Ohmic contacts and a superconducting electrode which is grounded (see Fig.~\ref{sample}) while the other In contact measures the 2DEG potential. To obtain $dV/dI(V)$ characteristics, we sweep the dc current through the interface from -4~nA to +4~nA.  This dc current is modulated by a low ac (4~pA, 2~Hz) component. We measure both the dc ($V$) and ac ($\sim dV/dI$) components of the 2DEG potential by using a dc voltmeter and a lock-in amplifier, respectively. The latter is equipped by a preamplifier with the 100~M$\Omega$ input impedance. We have checked, that the lock-in signal is independent of the modulation frequency in the 1~Hz -- 6~Hz range, which is defined by applied ac filters.

Because of the relatively low in-plane  2DEG resistance (about 1~k$\Omega$ at present 2DEG concentration and mobility), and the low resistance of the superconducting NbN electrode, the measured $dV/dI(V)$ curves reflect the behavior of the NbN-2DEG interface. To extract features specific to the HgTe edge transport, the measurements were performed at a temperature of 30~mK. Qualitatively similar results were obtained in several cooling cycles,  for different absolute values of the junctions'  $R_N$.


The examples of  $dV/dI(V)$ characteristics are presented in Fig.~\ref{IV} (a). They demonstrate a behavior, which is qualitatively consistent with a standard Andreev one for a single SN junction~\cite{tinkham}. A clearly visible NbN superconducting gap can be determined $\Delta_{NbN}=\pm$~1.58~mV (denoted by dashed lines in Fig.~\ref{IV}), which corresponds well to the directly measured critical temperature $T_c=11$~K for a similar NbN film. 

Within the gap at  $|eV|<\Delta_{NbN}$, the maximum differential resistance $R_{max}\approx 3$~M$\Omega$ is undoubtedly finite, which is only possible due to the Andreev reflection~\cite{tinkham}: if the Andreev process was suppressed, tunnel conductance would be zero at $|eV|<\Delta_{NbN}$. According to the BTK theory~\cite{tinkham}, a single-particle scattering is significant at the SN interface, because $R_{max}$ exceeds the normal junction resistance value $R_{N}\approx 0.5$~M$\Omega$. A corresponding transmission coefficient can be estimated~\cite{tinkham} as $R_N/R_{max}\approx 0.16$. Inset to Fig.~\ref{IV} (a) demonstrates, that  this ratio $R_N/R_{max}\approx 0.16$ is the same for the junctions with strongly different $R_N$.

On the other hand, the curves in the inset to Fig.~\ref{IV} (a) are characterized by  high $R_N$ values, 0.5~M$\Omega$ and 1.5~M$\Omega$ respectively. For a 20~$\mu$m wide junction and  $1.5 \cdot 10^{11}  $cm$^{-2}$ electron concentration, these $R_N$ correspond to the extremely low, below 10$^{-3}$, junction transmission. This should result~\cite{tinkham} in a pure tunnel regime with zero conductance at  $|eV|<\Delta_{NbN}$. This is obviously not the case in Fig.~\ref{IV} (a), so there is a clear contradiction between a well-developed Andreev behavior and high  $R_N$ values.  

Surprisingly, we also observe pronounced $dV/dI$ oscillations within the superconducting gap at  $|eV|<\Delta_{NbN}$ for the lowest (30~mK) temperature, see Fig.~\ref{IV} (a). The oscillating behavior is suppressed completely above 0.88~K, which is much below the critical $T_c=11$~K for our NbN film.

\begin{figure}
\includegraphics[width=\columnwidth]{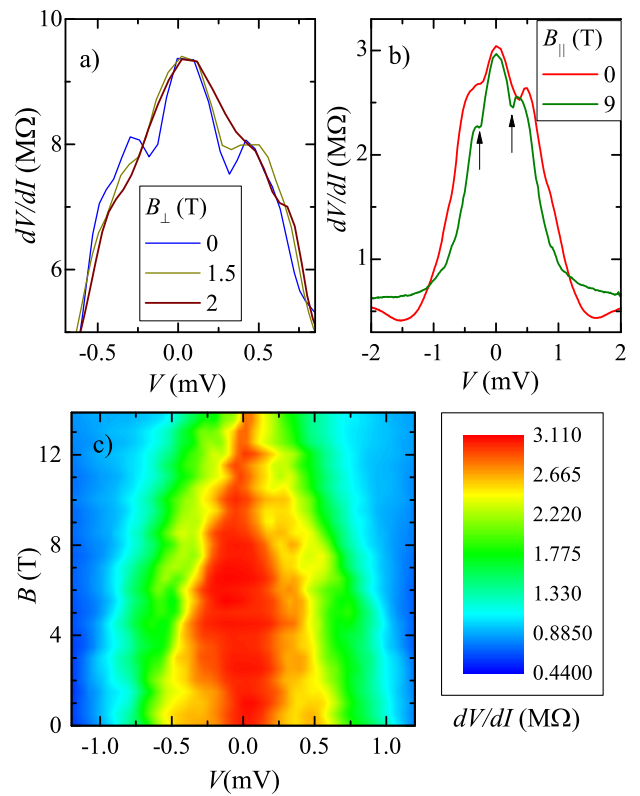}
\caption{(Color online) (a) Suppression of the conductance oscillations by the magnetic field, normal to the 2DEG plane. The oscillations disappear completely at $B_N=2$~T and do not reappear at higher fields. (b) Demonstration of the $dV/dI$ oscillating behavior for the in-plane field configuration, $B=9$~T. The oscillations' period is only weakly affected.  (c) Monotonous evolution of the $dV/dI(V)$ curve with in-plane magnetic field. The superconductivity is not suppressed up to 14~T because of a high critical field in NbN. All the curves are demonstrated for a minimal $T=30$~mK temperature. 
}
\label{ResMagn}
\end{figure}

The conductance oscillations can be shown in detail by subtracting the high-temperature (0.88~K) monotonous $dV/dI$ curve from the low-temperature (30~mK) oscillating one. The result is shown in Fig.~\ref{IV} (b) for the curves from the inset to Fig.~\ref{IV} (a), i.e. for two junctions with strongly different $R_N$ (0.5~M$\Omega$ and 1.5~M$\Omega$ respectively). The positions of the oscillations are denoted by arrows in Fig.~\ref{IV} (b). They coincide well and correspond to a constant period  $\Delta V=0.48$~mV, while the oscillations' visibility 
 is evidently higher for the highest $R_N=$1.5~M$\Omega$. The oscillations are concentrated strictly within the superconducting gap and  demonstrate  a $dV/dI$ resistance maximum at zero bias. 

The observed conductance oscillations are sensitive to magnetic field. They disappear completely above the 2~T magnetic field, normal to the 2DEG plane, see Fig.~\ref{ResMagn} (a). In contrast, the oscillations survive well for the in-plane field configuration, see Fig.~\ref{ResMagn} (b). The monotonous evolution of the $dV/dI(V)$ curve is shown in Fig.~\ref{ResMagn} (c). It can be seen that  the NbN superconductivity is not suppressed, while the gap is obviously diminished  at $B=14$~T. Above $B\approx 10$~T, the oscillations' period is also diminishing.


As a result, we observe (i) a well-developer Andreev behavior of $dV/dI(V)$ curves for high $R_N\sim 1$~M$\Omega$ values; (ii) pronounced conductance oscillations at  $|eV|<\Delta_{NbN}$ for the lowest (30~mK) temperature.

\begin{figure}
\includegraphics[width=\columnwidth]{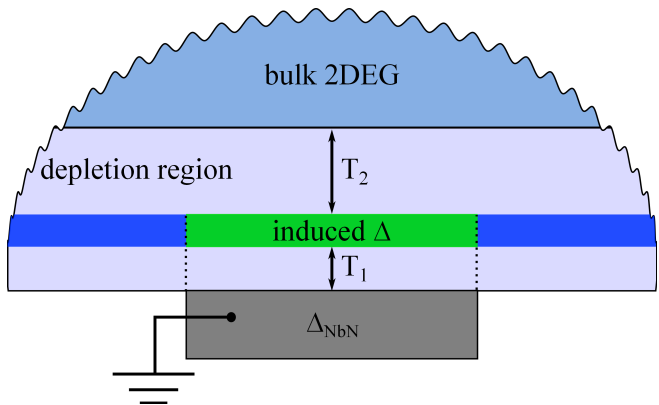}
\caption{(Color online) Top-view of the sample near the superconducting contact. A depletion region is shown, where the edge current is laterally localized. The transmission ${\cal T}_1\approx 0.16$, obtained from the Andreev curves, corresponds to  coupling between the edge conductive channel and the superconductive side NbN contact. Because of a proximity,   a gap $\Delta \sim {\cal T}_1\Delta_{NbN}<<\Delta_{NbN}$ is opened in a one-dimensional helical channel. Thus, the visinity of the NbN contact is equivalent to a one-dimensional NSN structure, as predicted in Ref.~\protect\onlinecite{adroguer}.
}
\label{discussion}
\end{figure}

Let us start the discussion from the high-temperature curve in Fig.~\ref{IV} (a). We observe a well-developed Andreev curve for high values of normal junction resistance $R_N$. This contradiction can be removed, if there is a conductive channel inside the insulating region.  In this case, $R_N$ is determined by the product $1/{\cal T}_1 {\cal T}_2$, where the transmission ${\cal T}_1\sim R_N/R_{max}\approx 0.16$ corresponds to high coupling between this conductive channel and the NbN side contact, while this channel is weakly (${\cal T}_2$ below 10$^{-2}$) coupled to a bulk 2DEG, see Fig.~\ref{discussion}.  

The edge conductive channel is well known in narrow HgTe quantum wells as a current-carrying helical edge state~\cite{konig}, if the well thickness exceeds the critical 6.3~nm~\cite{pankratov,zhang1,kane,zhang2}. The helical edge current has been demonstrated even to coexist with the conductive bulk by a direct visualization experiment~\cite{imaging}. This coexistence is also allowed by the theoretical considerations~\cite{pankratov,volkov}, but requires a low coupling between the edge channel and the bulk. This is possible~\cite{imaging} in samples with etched mesa, where a depletion region of finite width is always present at the 2DEG edge, because of the smooth edge potential~\cite{shklovskii,image02}. 

Thus, in a good accordance with previous investigations~\cite{konig,imaging}, we can expect that a well-developed Andreev curve is determined by transport between the NbN side contact  and the conductive helical edge channel. A corresponding ${\cal T}_1 \approx 0.16$ value is mostly determined by the mismatch in Fermi velocities in NbN and the helical channel, and therefore ${\cal T}_1$ is approximately the same for different junctions, as we can see from the perfect scaling of $dV/dI (V)$ curves in the inset to Fig.~\ref{IV} (a). In contrast, high values  $R_N \sim 1/{\cal T}_1{\cal T}_2$ are mostly determined by low ${\cal T}_2$, which reflects a low transmission of a  depletion region  at the 2DEG edge. The latter is sensitive~\cite{shklovskii,image02} to the long-range fluctuations of the edge potential, i.e. is expected to be different for different junctions.

At low temperatures, we observe a number of equally spaced conductance oscillations.  The oscillating behavior could be anticipated for a one-dimensional channel in a proximity with a superconductor due to either multiple Andreev reflections~\cite{tinkham,MAR} or Andreev bound states formation~\cite{chevallier}. The former leads to resonances~\cite{tinkham} at $E_n=2\Delta_{NbN}/n, n=1,2,3...$, which are not equally spaced~\cite{MAR}. The latter produces equally-spaced modulation of the density of states~\cite{chevallier}, but requires a short contact $L<\xi=\hbar V_F/\Delta$, where $\xi$ is a coherence length, $v_F$ is a Fermi velocity, $\Delta$ is a superconducting gap, induced in a one-dimensional channel.

The opposite limit of a long contact $L>>\xi$ has been considered in Ref.~\onlinecite{adroguer} for a helical edge channel, realized in a narrow HgTe quantum well. Due to a proximity with NbN, a gap $\Delta \sim {\cal T}_1\Delta_{NbN}<\Delta_{NbN}$ is opened in a one-dimensional  channel, see Fig.~\ref{discussion}. Thus, the channel is equivalent to a one-dimensional NSN structure in the vicinity of the NbN contact. Electrons with energies $E<\Delta$ experience total local Andreev reflection at each SN interface, because of the helicity conservation~\cite{adroguer}. Electrons of very high energy $E>>\Delta$ are perfectly transmitted.  For intermediate energies $E>\Delta$ Bogoliubov quasiparticles experience Fabry-Perot-type transmission resonances within the superconductive region.

In our experiment, the conductive edge channel is coupled to the NbN side contact,  so the resonances modulate the transmission ${\cal T}_1$. Thus, our experiment is somewhat equivalent to the described in Ref.~\onlinecite{finck}, where a three-dimensional topological insulator $Bi_2Se_3$ has been sandwiched between a superconducting and a normal lead, despite another limit $L<\xi$ in Ref.~\onlinecite{finck}, another dimension of a topological surface state, and different geometry of the experiment. The oscillations' period can be estimated~\cite{adroguer} as $ {\cal T}_1 e\Delta V \sim \pi\hbar v_F/L$. From the experimental period $\Delta V=0.48$~mV, contact width $L=20\mu$m, and ${\cal T}_1\approx 0.16$ we can estimate the Fermi velocity as $v_F\approx 8\times10^{7}$, which is in a reasonable agreement with the calculated~\cite{raichev} $v_F\approx 5\times10^{7}$. It's worth noting, that Andreev bound states formation would require a similar expression for the oscillations' period~\cite{chevallier}. However, we prefer the model of Ref.~\onlinecite{adroguer}, (i) because of the limit $L=20 \mu\mbox{m} > \xi\approx 2 \mu\mbox{m}$, and (ii) since the Andreev bound states should be primary seen in  a coherent transport, i.e. in a supercurrent.

The conductance oscillations, therefore, are connected only with the electrons, which are approaching the NbN contact along the helical channel. A higher oscillations' visibility for the highest $R_N=$1.5~M$\Omega$ in Fig.~\ref{IV} (b) can be well understood: higher $R_N$ indicates higher decoupling of the helical channel from the bulk (i.e. lower ${\cal T}_2$), which   increases the number of electrons which are approaching the NbN contact along the channel. On the other hand, the oscillations' period does not depend on ${\cal T}_2$, which  is consistent with the experimental observation of a constant period $\Delta V=0.48$~mV for junctions with strongly different $R_N$ in Fig.~\ref{IV} (b).  The oscillations' visibility is suppressed above $\Delta_{NbN}$, since the coupling to the NbN side contact is increased, see Fig.~\ref{IV}.

The proposed picture is also qualitatively consistent with the oscillations suppression at high temperatures and with the behavior in high magnetic fields. (i) The temperature smears the induced gap at $T \sim \Delta \sim {\cal T}_1\Delta_{NbN} \sim$~1~K, so a helical edge channel is normal in the vicinity of the NbN contact. This estimated temperature is consistent with the observed oscillations' disappearance at $0.88$~K in our experiment. (ii) A normal magnetic field induces Landau quantization in a 2DEG. One-dimensional transport is also allowed in this case~\cite{buttiker}, so the Andreev behavior of the $dV/dI (V)$ curve is not seriously changed in Fig.~\ref{ResMagn} (a). However, this edge transport is chiral~\cite{buttiker}, which forbids Fabry-Perot transmission resonance formation, proposed in   Ref.~\onlinecite{adroguer}. In our samples, a well-developed Landau quantization appears in 2~T, which is consistent with the field of the oscillations' disappearance. (iii) In-plane magnetic field does not destroy the edge channel helicity~\cite{raichev}. The superconducting gap is still resolved, see Fig.~\ref{ResMagn} (c), so the conductance oscillations survive in highest in-plane magnetic fields. According to the calculation~\cite{raichev}, $v_F$ is diminishing in high in-plane fields, which is also consistent with the observed diminishing of the  oscillations' period.

In conclusion, we experimentally investigate  electron transport through the interface between a superconductor and the edge of a two-dimensional electron system with band inversion. The interface is realized as a tunnel NbN side contact to a narrow 8~nm HgTe quantum well. It demonstrates a typical Andreev behavior with finite conductance within the superconducting gap. Surprisingly, the conductance is modulated by a number of equally-spaced oscillations. The oscillations are present only within the superconducting gap and at lowest, below 1~K, temperatures. The oscillations disappear completely in magnetic fields, normal to the two-dimensional electron system plane. In contrast,  the oscillations' period is only weakly affected by the highest, up to 14~T, in-plane oriented magnetic fields. We interpret this behavior as the predicted~\cite{adroguer} interference oscillations in a helical one-dimensional edge channel due to a proximity with a superconductor.


We wish to thank  A.M.~Bobkov, I.V.~Bobkova, Ya.~Fominov, V.T.~Dolgopolov, and T.M.~Klapwijk for fruitful discussions.  We gratefully acknowledge financial support by the RFBR (project No.~13-02-00065), RAS and the Ministry of Education and Science of the Russian Federation under Contract No. 14.B25.31.0007.

\end{document}